# Anomalous Platinum and Oxygen Transport during Electroforming of NbO$_x$ Memristors

Shimul Kanti Nath[1,2,5]*, Sanjoy Kumar Nandi[2]*, Xiao Sun[3], Sujan Kumar Das[2,6], Bin Gong[4], Nicholas J. Ekins-Daukes[5], Deepak Mishra[1], Mahesh P. Suryawanshi[5], William D. A. Rickard[3], Songyan Yin[4], Michael P. Nielsen[5] and Robert G. Elliman[2]*

[1]School of Electrical Engineering and Telecommunications, University of New South Wales (UNSW Sydney), Kensington NSW 2052, Australia

[2]Department of Electronic Materials Engineering, Research School of Physics, The Australian National University, Canberra ACT 2601, Australia

[3]John de Laeter Centre, Curtin University, Perth, WA 6102, Australia

[4]Solid State and Elemental Analysis Unit, Mark Wainwright Analytical Centre, Surface Analysis Laboratory, University of New South Wales (UNSW Sydney), Kensington NSW 2052, Australia

[5]School of Photovoltaic and Renewable Energy Engineering, University of New South Wales (UNSW Sydney), Kensington NSW 2052, Australia

[6]Department of Physics, Jahangirnagar University, Dhaka 1342, Bangladesh

*shimul_kanti.nath@unsw.edu.au

*sanjoy.nandi@anu.edu.au

*rob.elliman@anu.edu.au

# Abstract

Electroforming of metal-oxide-metal memristors is generally attributed to the creation of oxygen-vacancy filaments within the oxide, with noble metal electrodes such as Pt and Au remaining chemically inert. Here, we demonstrate that electroforming and subsequent operation of Pt/NbO$_x$/Nb$_2$O$_5$/Pt devices can induce an unexpected and highly correlated redistribution of both oxygen and platinum. Time-of-flight secondary ion mass spectrometry reveals a filamentary pathway characterized by micrometer-scale oxygen enrichment extending from the Nb$_2$O$_5$ layer through NbO$_x$ and deep into the Pt top electrode. Surprisingly, this is accompanied by the formation of a Pt-rich filament penetrating the oxide stack along the same filamentary path. Finite-element and lumped-element modelling show that current-controlled negative-differential-resistance operation produces localized Joule heating and high-frequency thermal cycling, which strongly enhances oxygen migration and enables thermally assisted Pt diffusion along vacancy-rich pathways. These findings reveal a previously unrecognized metal-ion transport mechanism in NbO$_x$ memristors and highlight the critical role of post-forming electrical dynamics in determining filament chemistry, stability, and device reliability.

## 1. Introduction

Memristive switching in two-terminal metal-oxide-metal (MOM) devices has attracted extensive research interest due to its promising applications in non-volatile memory technologies and neuromorphic computing [1-5]. When subjected to electrical stress, MOM devices typically exhibit non-volatile, volatile, or combined resistive switching responses. The observation of a particular response depends on the stoichiometry of the oxide film, device geometry (i.e., vertical or planar structure), choice of electrodes (i.e., reactivity and oxygen affinity of the electrode), and operating conditions (i.e., the magnitude of current or voltage stress).[6-12]

In most oxide-based memristors, an initial electroforming step is required to activate memristive switching by creating nanoscale conductive filaments that localize current flow between the electrodes [13-16]. This process is commonly attributed to electric-field-induced defect (oxygen vacancy) generation and drift and diffusion of oxygen vacancies. Associated Joule-heating can lead to further compositional and structural modification of the oxide and the metal/oxide interface, leading to a permanent filamentary volume whose composition and stability ultimately determine device performance and reliability [15, 17-21]. Therefore, a thorough understanding of electroforming and its microstructural consequences is vital for advancing memristive switching technologies and for accurate device modelling and optimization.

Among various oxide systems, niobium oxides ($NbO_x$) have garnered significant research interest due to their excellent thermal stability, reliable switching behaviour, and versatility in exhibiting different memristive responses [6, 22-26]. $NbO_x$ devices have found applications as selector elements in resistive random access memory (ReRAM), oscillatory neurons for neuromorphic circuits, and spike encoding components for machine vision [26-28].

Like many other oxides, $NbO_x$ devices typically require an electroforming step, mediated by either dielectric breakdown or current bifurcation processes [13]. Despite extensive electrical characterization, however, the microscopic evolution of conductive filaments during electroforming and subsequent device operation in $NbO_x$ remains incompletely understood, especially in multilayer $NbO_x/Nb_2O_5$ stacks where multiple oxygen reservoirs coexist.

Previous studies of oxide-based memristors have largely assumed that noble metal electrodes, such as platinum (Pt), remain chemically and structurally inert during switching, serving merely as electrical contacts. While several phenomena, such as oxygen exchange across metal/oxide interfaces, lateral migration of oxygen and Ta ions (in $TaO_x$), and local crystallization, have been widely reported for various oxide memristors [15, 18, 20, 29], but cationic migration of Pt into oxide layers has been generally considered negligible due to its low diffusivity and solubility. Additionally, comprehensive microstructural analyses of filaments in $NbO_x$ remain scarce, primarily due to the difficulty of detecting these nanoscale features in micrometre-scale devices widely investigated in the literature, as well as the significant variability among devices. Additionally, no prior study has provided direct 3D chemical imaging that correlates oxygen transport and metal ion migration. To date, most filament models neglected the transport of noble metals and focused exclusively on electric-field-driven oxygen vacancy redistribution and thermally driven compositional redistribution within the oxide (Soret Effect).

Here, we show that this assumption breaks down in electroformed $Pt/NbO_x/Nb_2O_5/Pt$ memristors operating in the negative differential resistance (NDR) regime. Using three-dimensional time-of-flight secondary ion mass spectrometry (ToF-SIMS), we directly visualize correlated filamentary transport of oxygen and platinum extending across the entire device stack. By combining these observations with finite-element and lumped-element electrothermal modeling, we demonstrate that post-forming electrical dynamics, particularly oscillatory current flow and associated thermal cycling, play a crucial role in driving both oxygen migration and thermally assisted Pt diffusion. These results reveal a previously unrecognized mechanism of metal-ion transport in $NbO_x$ memristors and underscore the importance of post-forming device behaviour in governing filament chemistry, microstructural evolution, and long-term reliability.

## 2. Experimental

Studies were conducted on 20 μm × 20 μm MOM cross-point devices comprising Pt/ $Nb_2O_5$Nb(O)/Pt structures grown on thermally oxidized Si substrates. The bottom metal electrodes consisted of a 25 nm Cr adhesion layer and a 40 nm Pt layer that were e-beam evaporated and patterned using a standard photolithographic lift-off process. A thin $Nb_2O_5$ film was then deposited by RF sputtering from an $Nb_2O_5$ target. Top electrodes were defined by a second photolithographic lift-off process and consisted of a Nb adhesion layer and a 25 nm Pt layer. We note that as a consequence of its reactivity and the relatively high base pressure, the Nb adhesion layer was partially oxidized during deposition to form a ~ 30 nm thick $NbO_{x\ (x\leq2)}$ layer. The final device structure was therefore 25 nm Pt (top electrode) / 30 nm $NbO_x$ / 27 nm $Nb_2O_5$/ 40 nm Pt (bottom electrode) / 25 nm Cr/300 nm $SiO_2$/Si. Further details of this structure is described in our earlier study[30].

Devices were tested using an Agilent B1500A semiconductor parameter analyzer attached to a Signatone probe station (S-1160). Bias sweeps were applied through the top electrodes, with the bottom electrodes grounded.

Devices were subsequently analysed using Scanning Electron Microscopy (SEM) (FEI Verios 460L) and ToF-SIMS (IONTOF Models M5 and M6). The latter employed a dual-beam depth profiling approach, with a 1 keV $Cs^+$ beam used for sputtering and a 30 keV $Bi^+$ primary ion beam used for analysis.

Finite Element modeling was performed using COMSOL Multiphysics and Lumped-element modeling was performed using LTSpice.

## 3. Results and Discussions

**3.1 Electroforming Behaviour and Localization of Conduction**

Fig. 1a shows a TEM image and EDX maps of a representative device cross-section, confirming the thickness and composition of the layered structure. The Pt/NbO$_x$/Nb$_2$O$_5$/Pt cross-point devices were initially in a high-resistance state and required electroforming to activate the NDR response. Fig. 1b shows a representative bidirectional current-controlled forming sweep, during which a sudden voltage drop occurs from ~6 V to ~1 V at an applied current of ~350 μA in the forward sweep (red line). In the reverse sweep (ranging from 800 μA to 0, as shown by the black line), the device exhibited an S-type NDR response. This behavior is consistent with the self-assembly of a localized threshold switching volume during the electroforming step[13] (further details of the electroforming process in NbO$_x$ films are given in the supporting information and also can be found in our earlier study[13]).

Following electroforming, scanning electron microscopy (SEM) reveals the appearance of a bright, circular feature approximately 1 μm in diameter on the top Pt electrode (Fig. 1c). Similar features have been reported in other oxide memristors [19] and are known to coincide with localized Joule heating and filamentary current flow. Previous in situ thermal mapping and resist-based filament detection techniques confirm that electrical conduction after forming in NbO$_x$ is strongly confined to these regions, identifying them as the active filamentary zones of the device [13, 31, 32].

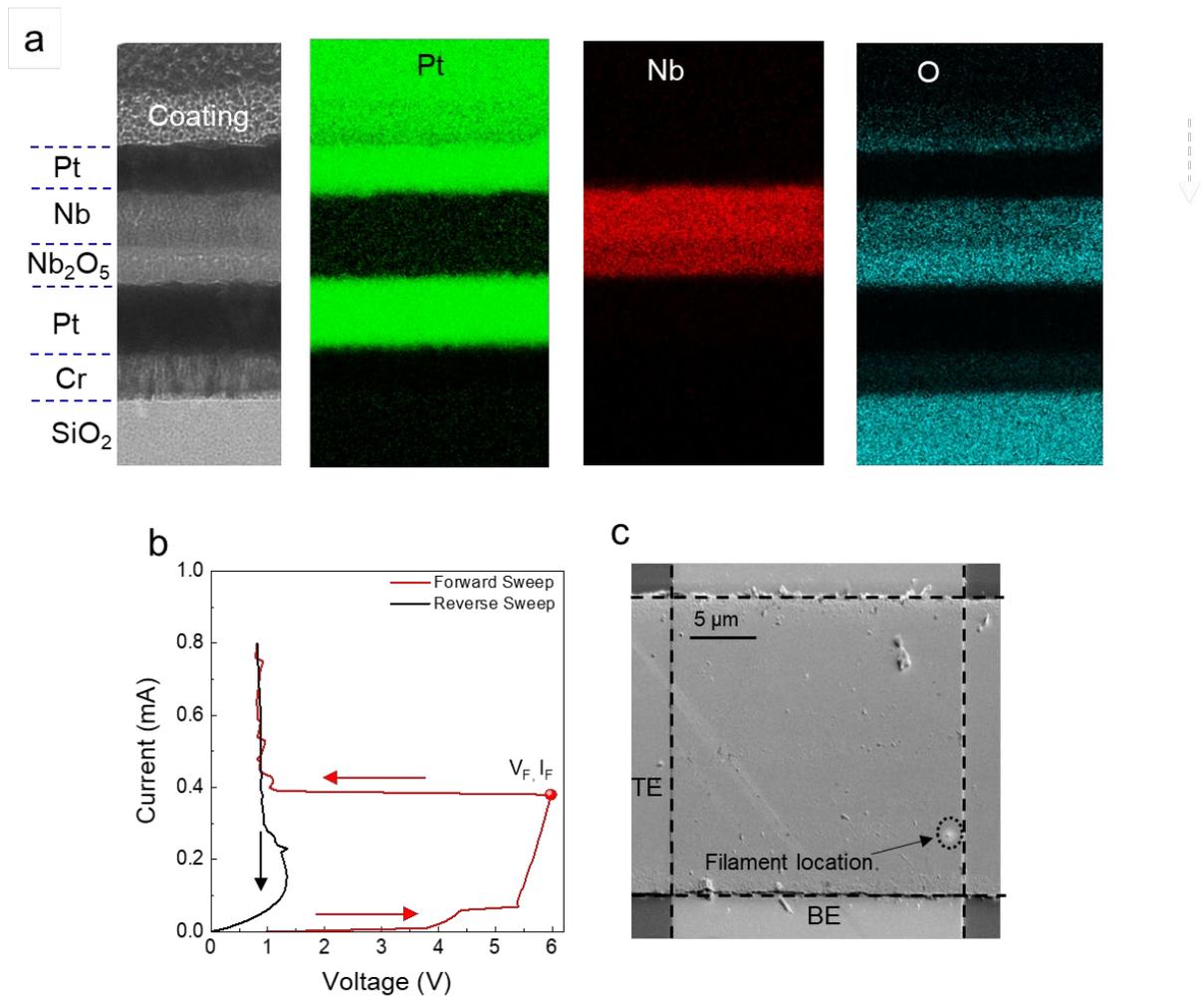

*Figure 1:* (a) TEM image and EDX maps of a device cross-section. (b) Electroforming step of a 20 μm device with Pt/Nb/Nb$_2$O$_5$/Pt structure under current-controlled testing with a bidirectional current sweep from 0 to 800 μA, and (c) SEM image of the top-view of the active device area (indicated by the dashed line) after electroforming, identifying filament location as a bright circular spot (TE and BE represent top and bottom electrodes, respectively).

### 3.2 Filament Composition Revealed by ToF-SIMS

In order to understand the compositional changes induced by electroforming, electroformed devices were examined using ToF-SIMS. The top panel of Fig. 2 shows reconstructed 3D concentration maps of Pt$^-$, O$^-$, Nb$^-$, NbO$^-$, and NbO$_2^-$ ions acquired under Cs$^+$ ion bombardment. A clear filamentary region is observed, spatially coincident across multiple ionic species and extending vertically through the oxide layers and into the top Pt electrode, as indicated by the dashed ellipses in the top panel. Lateral distributions of Pt and O within the various layers are shown in corresponding 2D maps (middle and bottom panels of Fig. 2), with the filamentary region highlighted by white circles.

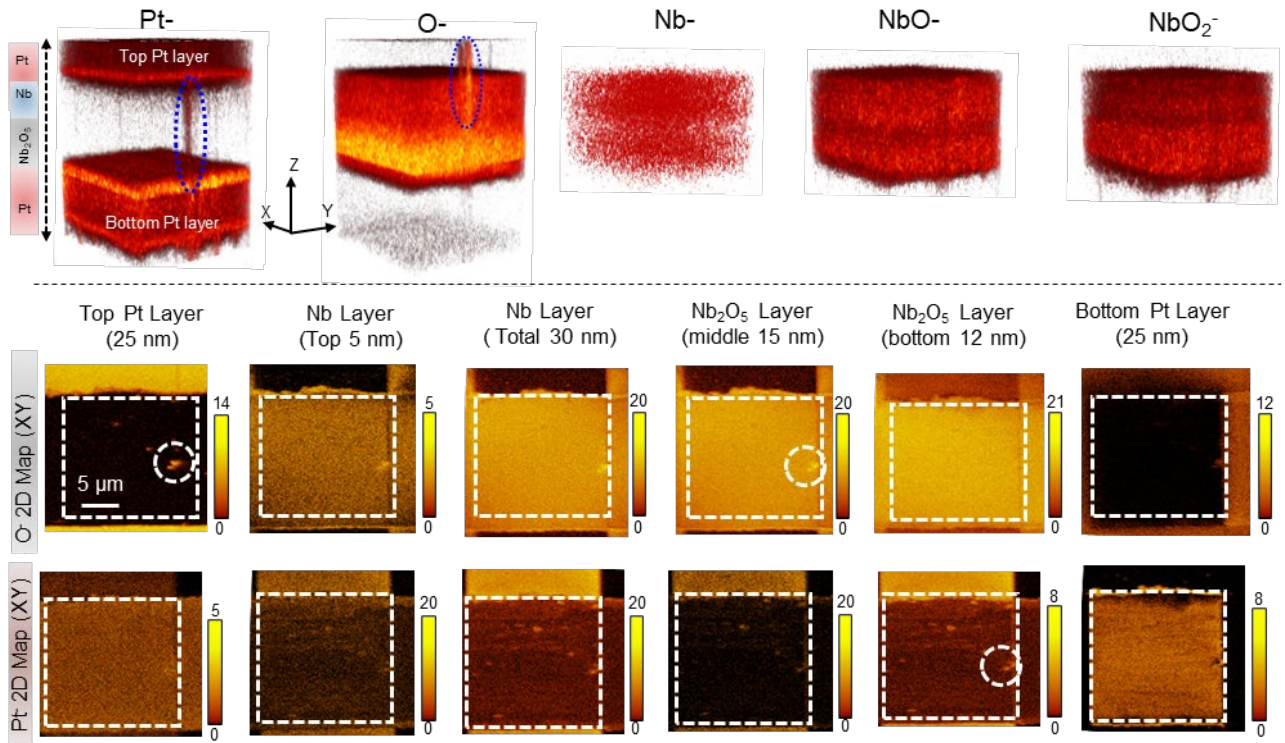

*Figure 2:* *ToF-SIMS images showing the distribution of different ions in a Pt/Nb(O)/Nb$_2$O$_5$/Pt device (maximum current applied through the device = 800 µA as shown in Fig. S1). The top panel shows 3D images of the selected device area with signals from Pt⁻, O⁻, Nb⁻, NbO⁻ and NbO$_2$⁻. The Pt- and O- 3D maps observe a distinct filamentary area, as marked by the dashed ellipses. The middle and bottom panels of Fig. 2 show 2D maps of O⁻ and Pt⁻ ions at different depths, showing the vertical and lateral extent of these ions as a function of layer thicknesses.*

These data reveal several interesting results. Most notably, the oxygen-rich filament is observed to extend through the NbO$_x$ and Pt layers, and Pt-rich filaments are observed to extend through the NbO$_x$ and Nb$_2$O$_5$ layers. The drift of oxygen towards the Pt anode, and its accumulation at the Pt/oxide interface are commonly observed during electroforming and is generally assumed to create an oxygen-deficient (i.e., oxygen vacancy) filament through the entire oxide film. Instead, these data show that oxygen originating from the stoichiometric $Nb_2O_5$ layer can be driven into the sub-stoichiometric $NbO_x$ layer, locally increasing its oxygen content. This indicates that post-formed conduction is dominated by a filamentary path primarily located within the $Nb_2O_5$ layer, with the $NbO_x$ layer acting as a dynamic oxygen reservoir rather than a purely vacancy-rich conductor. As shown in Fig. 2 (middle panel), the localized O distribution also extends through the Pt top-electrode layer.

Even more unexpectedly, the $Pt^-$ signal reveals a Pt-rich filament penetrating through both the $NbO_x$ and $Nb_2O_5$ layers along the same trajectory as the oxygen-rich filament. This observation is particularly surprising as Pt is generally regarded as an inert electrode material that exhibits minimal cationic drift under typical MOM device conditions and has very low solubility and diffusivity, which severely limits its mobility within the oxide matrix. The lateral extent of both oxygen and platinum filaments (~ 1 µm), observed in two-dimensional ion maps (middle and bottom panels of Fig. 2), closely matches the bright regions identified by SEM, further confirming that these compositional redistributions are confined to the electrically active filament region (further compositional analysis of the filamentary area by transmission electron microscopy (TEM) is given in supporting information, Fig. S3).

**3.3 Oxygen and Platinum Transport: Effect of Electroforming**

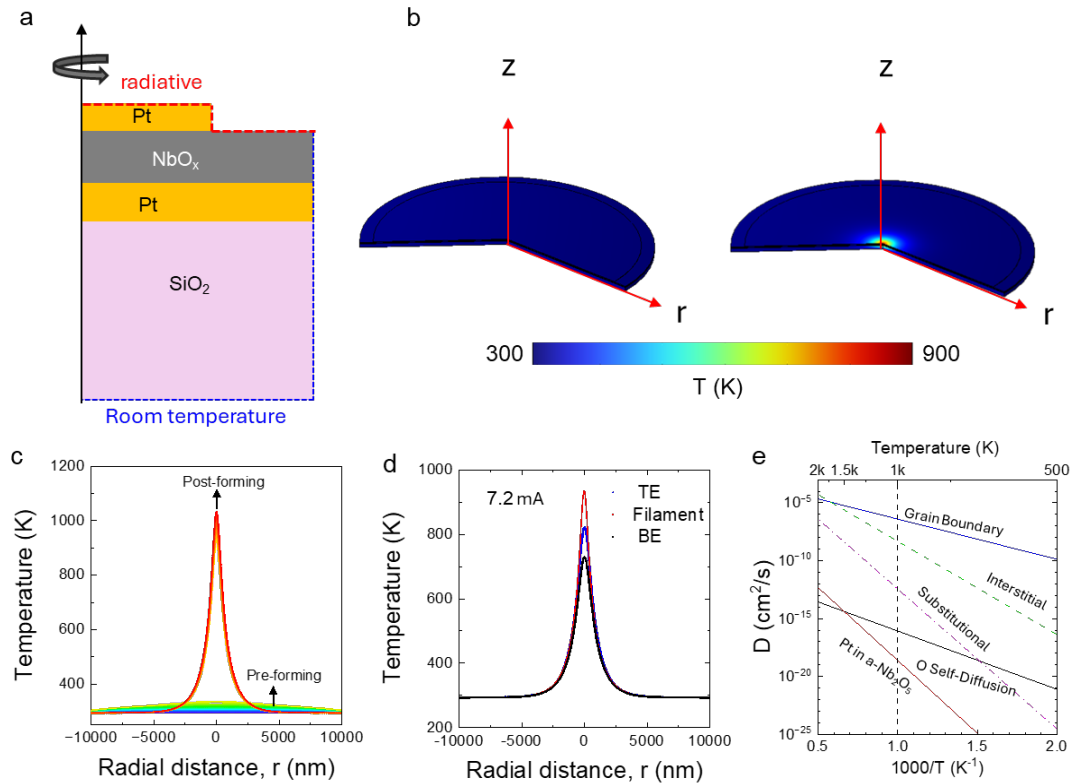

***Figure 3:*** *(a) Schematic of the device structure used in the COMSOL model. (b) 3D maps of temperature distribution within the device area before (left) and after filament formation (right), (c-d) radial temperature distribution after forming. (e) Diffusion coefficient (D) of O and Pt as a function of temperature calculated from Arrhenius type diffusion equation.*

To quantify the role of Joule heating on the transport of O and Pt during electroforming, we developed a two-dimensional axisymmetric finite-element model of a Pt/NbO$_x$/Pt/SiO$_2$ structure (Fig. 3a) that self-consistently solves the coupled current-continuity and heat-transfer equations. The model incorporates Poole-Frenkel conduction and Joule heating[13, 33] but does not assume any pre-existing material inhomogeneity. The simulations reproduce the experimentally observed voltage jump during electroforming (see supporting information, Fig. S5) and predict a spontaneous bifurcation of the current into a high-current-density filament at the device center. Fig. 3b shows 3D maps of temperature distribution within the simulated device area before (left panel) and after filament formation (right panel). It is evident that electroforming produces a sharp local temperature rise within the filament, accompanied by strong lateral thermal gradients (Fig. 3c-d) and redistribution of current density. These results are consistent with filamentary conduction observed experimentally and establish a direct link between electrical instabilities and localized self-heating during electroforming.

Using these temperature distributions as a guide, it is possible to estimate the extent of thermal diffusion for O and Pt. For this purpose, we assume that diffusion follows Arrhenius behaviour, expressed as $D = D_0 \exp(-E_a/k_B T)$, where $D$ is the diffusion coefficient, $D_0$ is the pre-exponential factor, $E_a$ the activation energy, $k_B$ the Boltzmann constant, and $T$ the absolute temperature. Reported diffusion parameters for oxygen and platinum diffusion are summarized in Table 1, and the corresponding diffusion coefficients are plotted in Fig. 3e.

**Table 1:** *Diffusion parameters for O in Pt and a-Nb$_2$O$_5$, and Pt in a-Nb$_2$O$_5$.*

| Species: Substrate | Mode | $D_0$ (cm²/s) | $E_a$ (eV) | D at 1000 K (cm²/s) | Diffusion length (1s at 1000K) (nm) | Ref. |
|---|---|---|---|---|---|---|
| **O: Pt** | Substitutional | 0.46 | 2.4 | 3.7×10⁻¹³ | 6 | 34 |
| | Interstitial | 0.56 | 1.6 | 4.8×10⁻⁹ | 693 | |
| | Grain Boundary | 1.2×10⁻³ | 0.69 | 4.0×10⁻⁷ | 6324 | |
| **O: a-Nb$_2$O$_5$** | Collective | 9.4×10⁻¹² | 1.0 | 8.6×10⁻¹⁷ | 0.09 | 35 |
| **Pt: a-Nb$_2$O$_5$** | Substitutional | 1×10⁻⁶ | 2.5 | 2.5×10⁻¹⁹ | 5.0×10⁻³ | |

Oxygen diffusion in a-Nb$_2$O$_5$ occurs via thermally activated, vacancy-mediated hopping within a disordered oxygen sublattice, strongly influenced by local structural heterogeneity. The absence of a periodic lattice leads to a broad distribution of migration barriers, with diffusion proceeding along low-energy percolation pathways involving bond breaking, bond switching, and cooperative network relaxation [35,36]. As oxygen vacancies act as the key transport enablers, diffusion is extremely slow in the temperature range of interest (see Table 1). For example, at 1000 K, the diffusion coefficient is ~$D \approx 8.6 \times 10^{-17}$ cm$^2$/s, corresponding to a diffusion length of ~0.09 nm after 1 s. The observed O redistribution must therefore be dominated by drift, which for applied electric fields in the range $10^8$-$10^9$ V/m can enhance O transport by $10^3$-$10^4$ times.

In contrast, O diffusion within the polycrystalline Pt electrodes is relatively fast, with grain boundaries providing both rapid diffusion pathways and energetically favourable trapping sites for oxygen (see Table 1). Interstitial diffusion is also relatively fast, but transport is limited in this case by the low solubility of O in Pt and by O-trapping at Pt vacancies. Grain boundary diffusion is therefore expected to dominate and is more than adequate to account for the transport of O through the Pt electrode layers. Enhanced trapping within the grain boundaries can also account for O retention within the Pt layer (see Figs. 1c and 2). Importantly, ToF-SIMS measurements show no evidence for the formation of Pt oxides.

The formation of Pt-rich filaments within the oxide is far less intuitive. Our finite-element simulations indicate that the peak filament temperature during post-forming operation reaches ~1000 K, which is insufficient to account for significant Pt diffusion within Nb$_2$O$_5$ (See Table 1 and Fig. 3c). Moreover, direct cationic drift of Pt under moderate electric fields is highly unlikely due to its large atomic mass and high activation energy for diffusion.

Consequently, Pt transport likely requires either localized redox processes that substantially reduce the effective diffusion barrier, such as oxygen-vacancy-assisted diffusion, or temperatures well above 1000 K. The latter can result from the discharge of parasitic and device capacitances during electroforming[30] and electrical oscillations during reverse current sweeps, where the NDR response of the device creates a relaxation oscillator. Such overshoot can produce large temperature excursions and may partially account for the anomalous Pt redistribution. To investigate these effects, we employed a lumped element electrothermal model implemented in LTspice, incorporating a filamentary conduction element, a parallel leakage resistance, and parasitic capacitance.

### 3.4 Oxygen and Platinum Transport: Effect of Transients

As noted above, the presence of a parallel capacitance across the device can have two important consequences: (i) discharge through the device during electroforming, producing large transient current spikes, and (ii) transformation of a device NDR into a relaxation oscillator. The latter is particularly significant, as it can induce large, periodic temperature excursions that continue to modify the material structure during the self-assembly of the threshold switching volume (during the forward sweep in Fig. 1b) and also after the initial electroforming step (during the reverse sweep in Fig. 1b).

To explore these effects, we simulated the current-voltage characteristics and filament temperature using a lumped-element circuit model. Filament conduction was described using a Poole-Frenkel-based electrothermal model [37], and embedded within a measurement circuit (Fig. 4a) that included a parallel resistance to account for non-filamentary conduction and a parallel capacitor representing device and instrumentation capacitances. Simulations were performed in LTspice for bidirectional current sweeps with parallel capacitances of 0-100 pF.

Model parameters were chosen to reproduce the reverse current sweep in Fig. 1b. Details of the thermal model and fitting parameters are provided in the supporting information.

The simulation results, summarized in Fig. 4, highlight the critical role of parallel capacitance. For capacitances of 0 and 20 pF, the device exhibits smooth NDR behaviour, whereas voltage oscillations emerge at higher capacitances (Fig. 4d). The oscillation window increases with capacitance and becomes asymmetric between forward and reverse sweeps for capacitances of 30-100 pF.

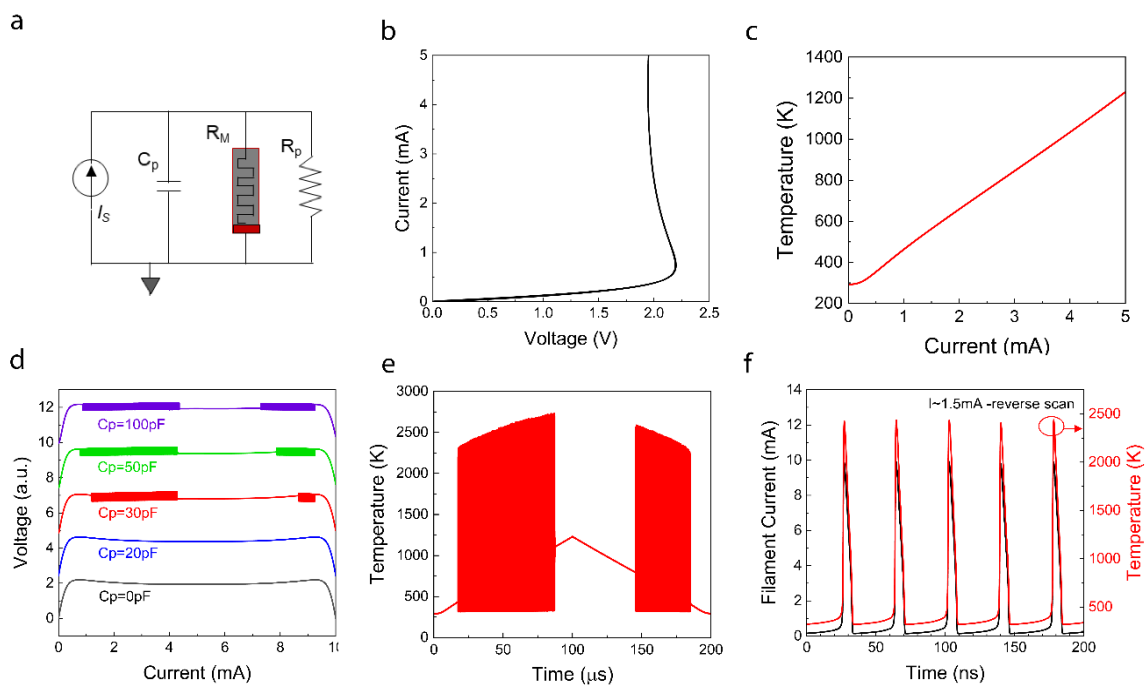

*Figure 4: (a) Schematic of the measurement circuit for LTspice modeling, (b) current-controlled I-V characteristic of a device with $C_p$=20 pF, and (c) associated average filament temperature. (d) Current-controlled I-V characteristic of a device with varying $C_p$= 0 to 100 pF, (e) filament temperature when $C_p$=30 pF, and (f) magnified view of filament current and temperature when $C_p$=30 pF, which clearly indicates the periodic oscillation of the device temperature and current.*

The simulated NDR response (Fig. 4b) shows a threshold voltage of approximately 2.25 V, consistent with that observed in our devices after electroforming. The peak filament temperature reaches ~900 K near 3 mA, gradually increases to around 1200 K at the end of the forward sweep (5 mA), and then decreases during the reverse sweep (Fig. 4c). This thermal regime provides a natural explanation for the ToF-SIMS observations of oxygen transport. Given the relatively high grain-boundary diffusivity of oxygen in Pt (~ $9.3 \times 10^{-8}$ cm² s$^{-1}$ at 900 K[34]) and a characteristic dwell time of ~ 0.1 s, oxygen redistribution within the Pt electrode is readily achieved, even allowing for some temperature drop within the metal. However, this temperature remains insufficient to drive measurable Pt diffusion in the oxide.

In contrast, the oscillatory response produces extreme and repetitive thermal stress that can enable thermally assisted Pt migration along O-vacancy-rich filament cores. Fig. 4d-f show simulation results for a device with a parallel capacitance of 30 pF, which operates as a relaxation oscillator with well-defined oscillation windows during both forward and reverse current sweeps. These oscillations induce large temperature excursions exceeding 2500 K, and although each excursion is short-lived (~ 0.2-10 ns), they recur at high frequencies (~ 3-14 MHz), subjecting the filament to repeated high-temperature cycling.

We therefore propose that Pt migration is enabled by a combination of (i) vacancy-assisted diffusion, in which oxygen vacancy fluxes locally reduce diffusion barriers in the filamentary region, and (ii) extreme, transient heating within the filament during oscillatory operation. In this scenario, oxygen vacancy accumulation creates low-energy diffusion pathways, while repeated thermal excursions activate Pt diffusion along these paths. This regime accounts for the enhanced redistribution of both oxygen and platinum observed experimentally (Fig. 2) and is also expected to play a dominant role in the local crystallization of the $Nb_2O_5$ and $NbO_x$ layers.

In this context, we note that we performed ToF-SIMS and TEM analyses on Pt/Nb/NbO$_x$/Pt structures with a single sputtered sub-stochiometric NbO$_x$ layer and observed similar Oxygen and Platinum migration as well as local crystallization (see supporting information), confirming the generality of the above results.

**Conclusions**

In summary, we demonstrate that electroforming and operation of Pt/NbO$_x$/Nb$_2$O$_5$/Pt memristors produce complex, strongly coupled ionic and metallic transport phenomena that extend well beyond the conventional oxygen-vacancy filament model. Three-dimensional ToF-SIMS provides the first direct experimental evidence of filamentary redistribution of oxygen from Nb$_2$O$_5$ into NbO$_x$ and deep into the Pt electrodes, accompanied by the unexpected formation of Pt-rich filaments penetrating the oxide stack along the same conductive pathways. By combining finite-element simulations with lumped-element electrothermal modeling, we have shown that oscillatory current flow associated with negative differential resistance generates extreme, repetitive thermal excursions that dramatically enhance both oxygen migration and thermally assisted platinum diffusion along vacancy-rich filaments. Based on these observations, we introduced a vacancy-assisted, thermally activated mechanism for Pt transport in NbO$_x$-based memristors, in which oscillatory NDR-driven heating enables noble-metal migration along defect-rich filaments formed during electroforming. More broadly, these findings demonstrate that Pt electrodes cannot always be regarded as inert in oxide memristors and highlight post-forming electrothermal instability as a key factor governing filament chemistry, structural evolution, and long-term device reliability in NbO$_x$-based and related memristive systems.

**Acknowledgments**

We acknowledge access to NCRIS facilities at the ACT nodes of the Australian National Fabrication Facility (ANFF) and Microscopy Australia, and the UNSW Mark Wainright Analytical Centre (MWAC). S.K.N. acknowledges the support from the ARC DP220101532. Xiao Sun acknowledges the support of John de Laeter Centre, Curtin University, Curtin Faculty of SAE R&DC Small Grants, and the ARC LIEF (LE190100053) grants. R.G.E. acknowledges support from ARC DP230100462. M.P.N. recognizes the support of the University of New South Wales Scientia Program and an Australian Research Council (ARC) Discovery Early Career Researcher Award (DECRA) Fellowship (Grant No. DE230100382).

**Unexpected Platinum Migration Accompanying Oxygen Transport in Electroformed NbO$_x$ Memristors**

Shimul Kanti Nath[1,2,5]*, Sanjoy Kumar Nandi[2]*, Xiao Sun[3], Sujan Kumar Das[2,6], Bin Gong[4], Nicholas J. Ekins-Daukes[5], Deepak Mishra[1], Mahesh P. Suryawanshi[5], William D. A. Rickard[3], Songyan Yin[4], Michael P. Nielsen[5] and Robert G. Elliman[2]*

[1]School of Electrical Engineering and Telecommunications, University of New South Wales (UNSW Sydney), Kensington NSW 2052, Australia

[2]Department of Electronic Materials Engineering, Research School of Physics, The Australian National University, Canberra ACT 2601, Australia

[3]John de Laeter Centre, Curtin University, Perth, WA 6102, Australia

[4]Solid State and Elemental Analysis Unit, Mark Wainwright Analytical Centre, Surface Analysis Laboratory, University of New South Wales (UNSW Sydney), Kensington NSW 2052, Australia

[5]School of Photovoltaic and Renewable Energy Engineering, University of New South Wales (UNSW Sydney), Kensington NSW 2052, Australia

[6]Department of Physics, Jahangirnagar University, Dhaka 1342, Bangladesh

*shimul_kanti.nath@unsw.edu.au

*sanjoy.nandi@anu.edu.au

*rob.elliman@anu.edu.au

1. **Electroforming**

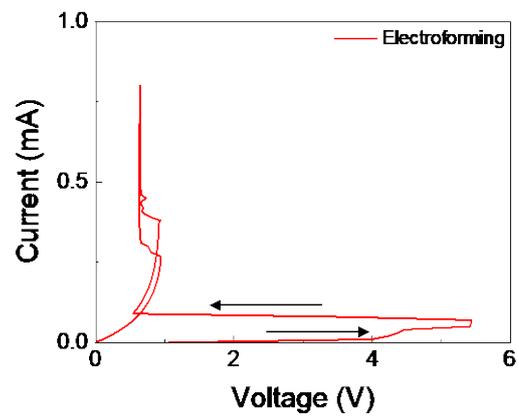

**Figure S1:** Bidirectional current sweep showing the electroforming step followed by an immediate NDR response in a 20 μm cross-point device that was further analyzed using ToF-SIMS (Fig. 3).

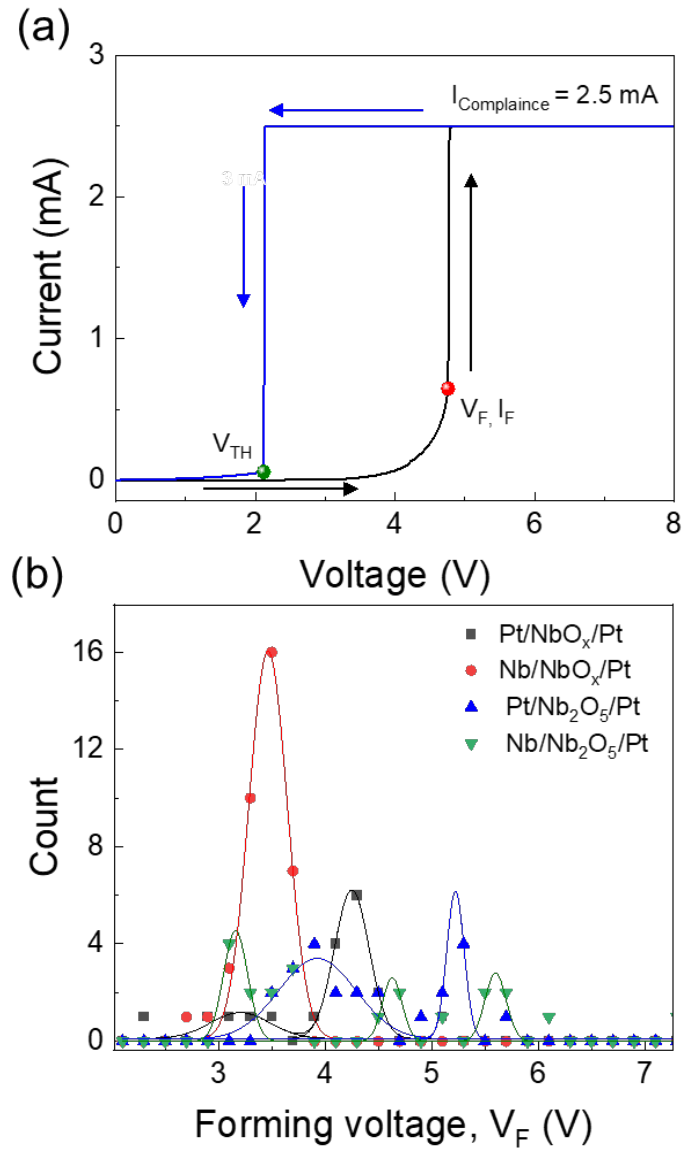

**Figure S2:** (a) Bidirectional voltage sweep showing a representative electroforming step (black line) followed by an immediate threshold switching response (blue line) under voltage-controlled testing in a 10 μm cross-point device with Pt (25 nm)/Nb (15 nm)/Nb$_2$O$_5$ (22 nm)/Pt (25 nm) structures. (b) Forming Voltage ($V_F$) distribution as a function of oxide stoichiometry and device structure.

## 2. Structural and compositional analyses of the filamentary area in Pt/Nb/Nb$_2$O$_5$/Pt structures revealed by SEM, TEM, and EDX.

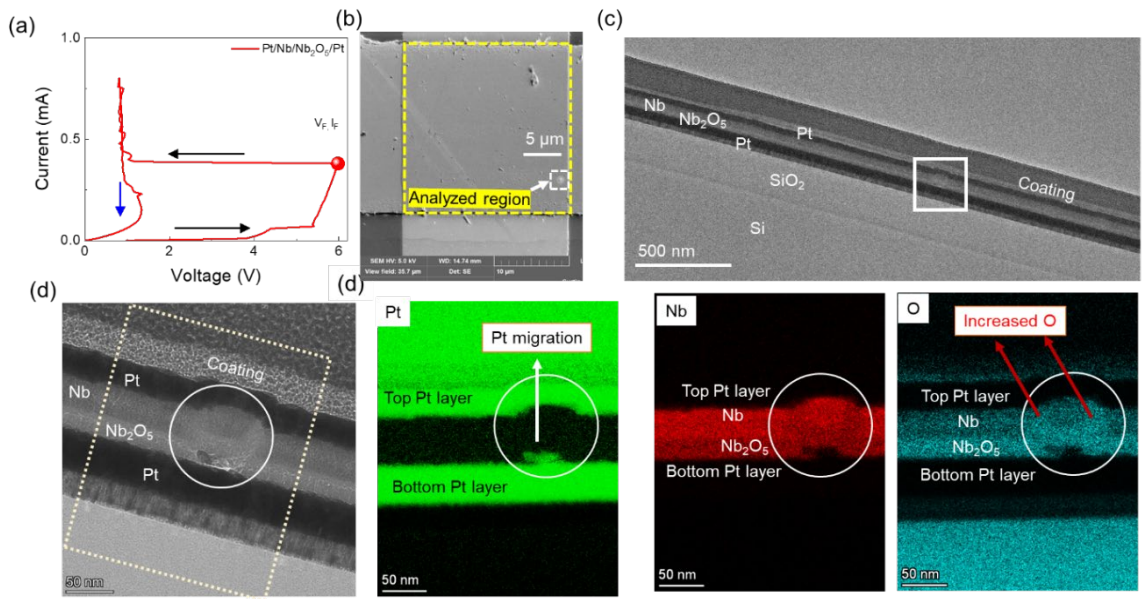

**Figure S3:** (a) Electroforming step of a 20 μm device with Pt/Nb/Nb$_2$O$_5$/Pt structure under current-controlled testing with a bidirectional current sweep from 0 to 800μA. (b) SEM image of the top-view of active device area (indicated by the dashed yellow line) after electroforming, identifying a bright circular spot. (c) TEM micrograph of the device area within the middle of the bright spot observed under SEM. (d) TEM micrograph showing the magnified view of the device area marked by the rectangle in Fig. S3(c), indicating nano-scale modification in the device structure. (e): EDX maps showing elemental distribution within the dashed rectangular area in Fig. S3(c-d). Note that, the TEM lamella size is around 100 nm, while the diameter of the filamentary area is around 1-2 μm in our device; therefore, TEM analysis is insufficient to capture the detailed information of the filament structure in a μm-scale device. ToF-SIMS analysis can provide more detailed information as discussed in the manuscript.

## 3. ToF-SIMS analysis of post-electroforming devices with sub-stoichiometric $NbO_x/Nb_2O_5$ films.

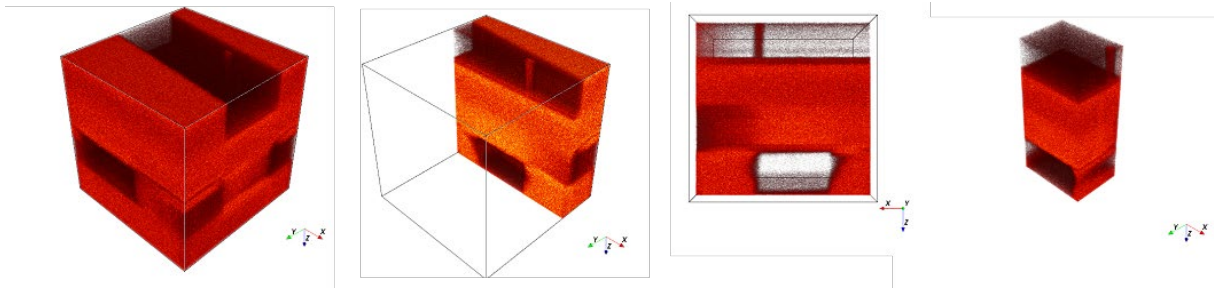

**Figure S4:** (a) ToF-SIMS images showing oxygen ($O^-$) distribution in a Pt/Nb/Nb$_2$O$_5$/Pt device (maximum current applied through the device = 800 µA). The selected device area was cropped and rotated to visualize the extent of the filamentary area.

## 4. COMSOL Parameters

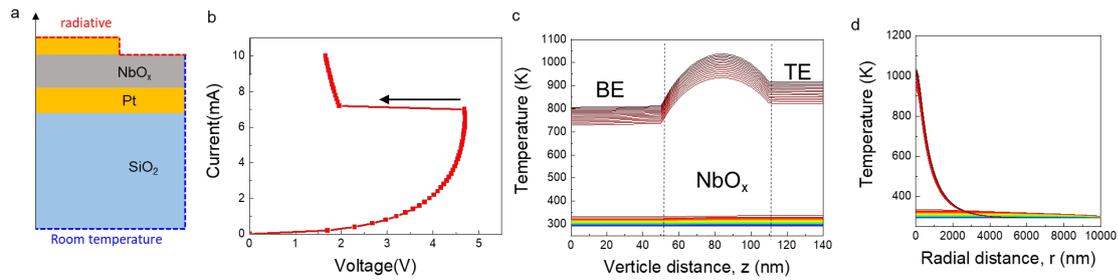

**Figure S5:** (a) Schematic of the device structure (b) Simulated I-V curve (c-d) temperature in different layers as a function of vertical and radial distance, respectively.

The finite element simulations employed the electric currents and heat transfer modules in COMSOL to simultaneously solve the coupled electrical and thermal transport equations.

Thermal boundary conditions included radiative heat loss from the top surface of the device, with an emissivity value of 0.02. The outer radial boundary of the device and the bottom surface of the SiO$_2$ layer were maintained at a constant ambient temperature of 293 K. Electrical simulations were performed under current-controlled conditions, where the current source was applied to the top Pt electrode while the bottom Pt electrode was grounded.

The material parameters for Pt and SiO$_2$ were obtained from the COMSOL material database. Electrical transport in the NbO$_x$ layer was assumed to follow Poole-Frenkel conduction, described by the following equation.

$$\sigma = \sigma_0 \exp\left(-\frac{E_a + \beta\sqrt{E}}{kT}\right)$$

Where, $\beta = \frac{e^3}{\pi \varepsilon_0 \varepsilon_r}$

The relevant parameters used for NbOx are listed in Table S1.

Table S1: Model parameters used in finite element model

| Parameter | Value | Unit |
| --- | --- | --- |
| Bottom Electrode | 50 | nm |
| Top Electrode | 25 | nm |
| Oxide thickness | 60 | nm |
| $\sigma_0$ | $1\times10^5$ | S/m |
| $k_{th}$ | 1 | W/mK |

5. **LTSPICE Parameters**

In the LTSpice model, we considered a threshold-switching memristor in which electrical conduction is assumed to follow a Poole-Frenkel transport mechanism. Under this assumption, the resistance of the memristive region can be expressed as[37]:

$$R_m = R_0 \exp\left(\frac{E_a - q\sqrt{\frac{qE}{\pi \varepsilon_0 \varepsilon_r}}}{k_B T}\right)$$

where $k_B$ is the Boltzmann constant, $E_a$ represents the activation energy for conduction, $\varepsilon_0$ is the vacuum permittivity, and $\varepsilon_r$ is the relative permittivity of the threshold-switching region. $T_m$ denotes the temperature of the electrically active region, while $T_{amb}$ represents the ambient temperature. $R_0$ is the resistance prefactor corresponding to the active region at $T = T_{amb}$.

The time-dependent thermal behaviour of the device is described using Newton's law of cooling:

$$\frac{dT_m}{dt} = \frac{I_m^2 R_m}{C_{th}} - \frac{\Delta T}{R_{th} C_{th}}$$

where $R_{th}$ and $C_{th}$ represent the thermal resistance and thermal capacitance of the device, respectively. The term $\Delta T$ corresponds to the temperature difference between the active region ($T_m$) and the ambient environment ($T_{amb}$). The parameters used in the model are summarized in Table S2.

Table S2: Model parameters used in lumped element model

| Parameter | Value | Unit |
| --- | --- | --- |
| Oxide thickness | 50 | nm |
| $E_A$ | 1.0 | eV |
| $R_0$ | 120 | Ω |
| $C_{th}$ | $1\times10^{-15}$ | J/K |
| $R_{th}$ | $1\times10^5$ | K/W |

## 6. ToF-SIMS analysis of post-electroforming devices with sub-stoichiometric NbO$_x$ films.

To explore the nature of ion migration in sub-stoichiometric NbO$_x$ films, additional measurements were carried out on devices with Pt/Nb (5 nm)/NbO$_x$/Pt structures, which predominantly exhibited threshold switching or NDR behaviour after electroforming. In all cases, the same compositional trends were observed: oxygen exchange at the top Pt electrode and Pt migration from the bottom electrode toward the top, as illustrated in Fig. 4c. It should be noted that the lateral resolution of the ToF-SIMS system (~50 nm) limits the ability to resolve nanoscale lateral ion movements. As a result, subtle thermally driven ion redistribution effects, such as those arising from temperature gradients (known as the Soret effect) and previously reported in TaO$_x$-based devices, [17, 38] could not be directly resolved in the present measurements.

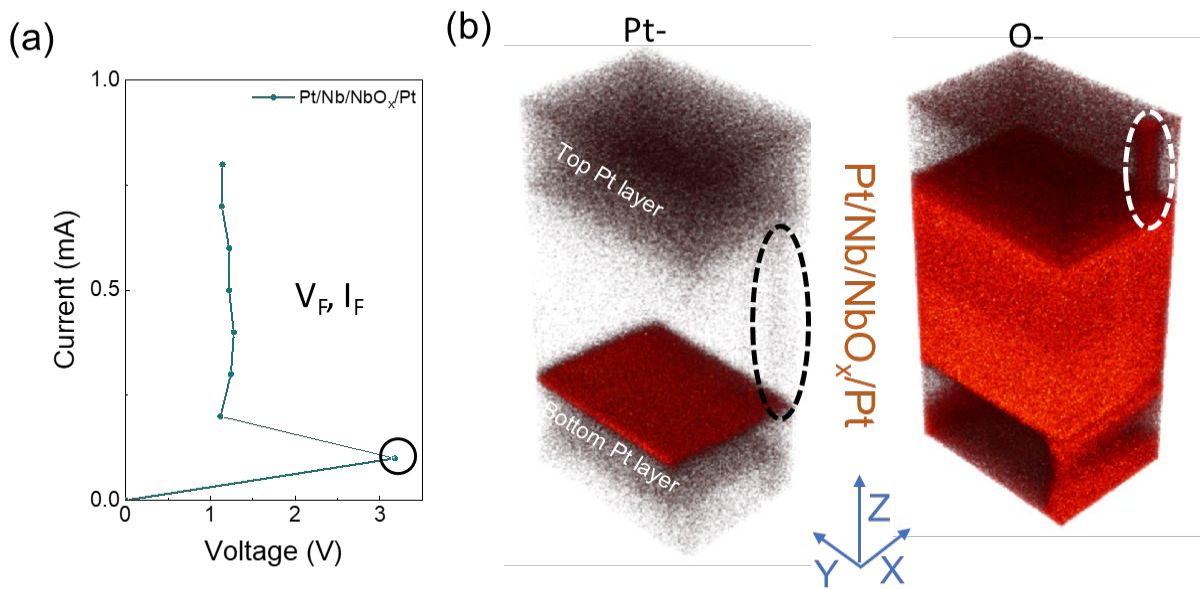

**Figure S6:** (a) Current-controlled Electroforming (unidirectional sweep) and (b) 3D ToF-SIMS maps of a Pt/Nb (5 nm)/NbO$_x$/Pt device with signals from Pt$^-$, O$^-$.

## 7. Structural and compositional analyses of the filamentary area under extreme biasing conditions and repeated cycling.

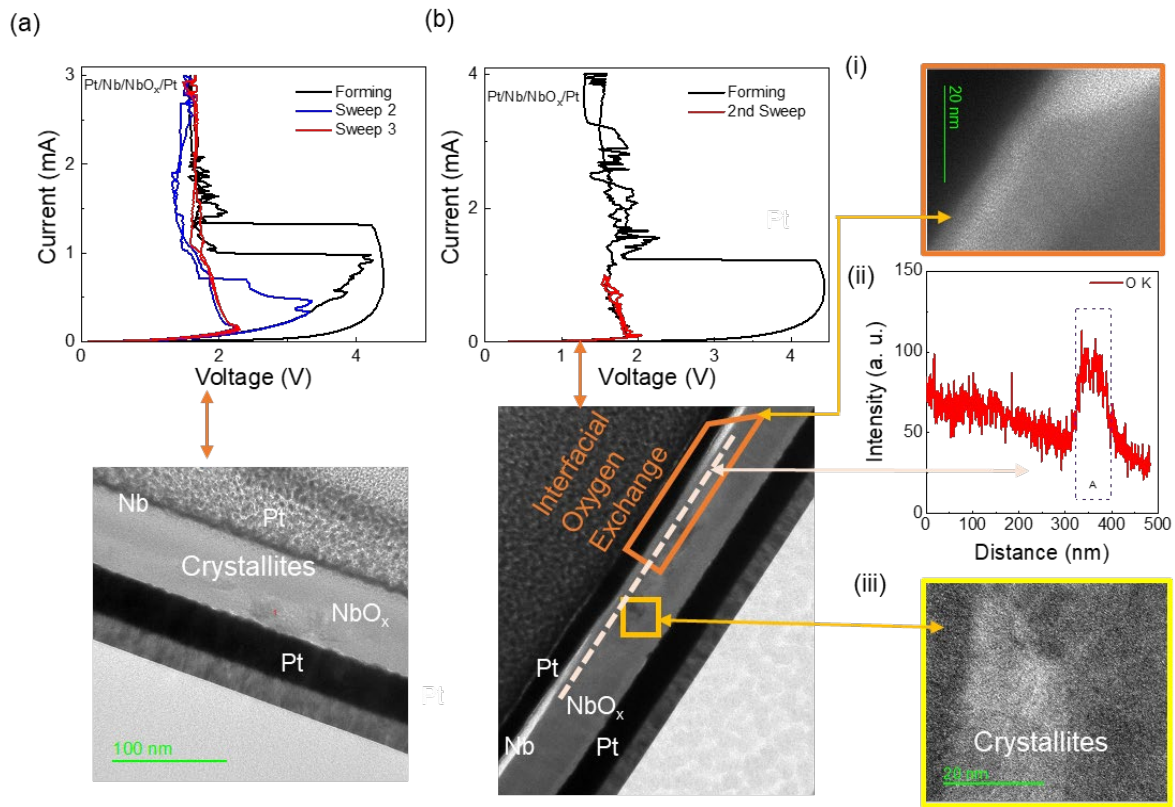

**Figure S7:** Electroforming and TEM imaging of the filamentary area in devices with (a) 10 µm Pt/Nb(5 nm)/NbO$_x$/Pt and (b) 10 µm Pt/Nb(10 nm)/NbO$_x$/Pt structures. Inset (b-i) shows an enlarged view of the selected device area showing contrast near the top Pt electrode, (b-ii) EDX map obtained from a line scan through NbO$_x$ layer close to the top electrode/oxide interface (scan direction is shown inset by the dashed line in Fig. (db). Inset (b-iii) shows an enlarged view of the marked area in Fig. (b).

## 8. Temperature-dependent I-V characteristics of a post-forming device

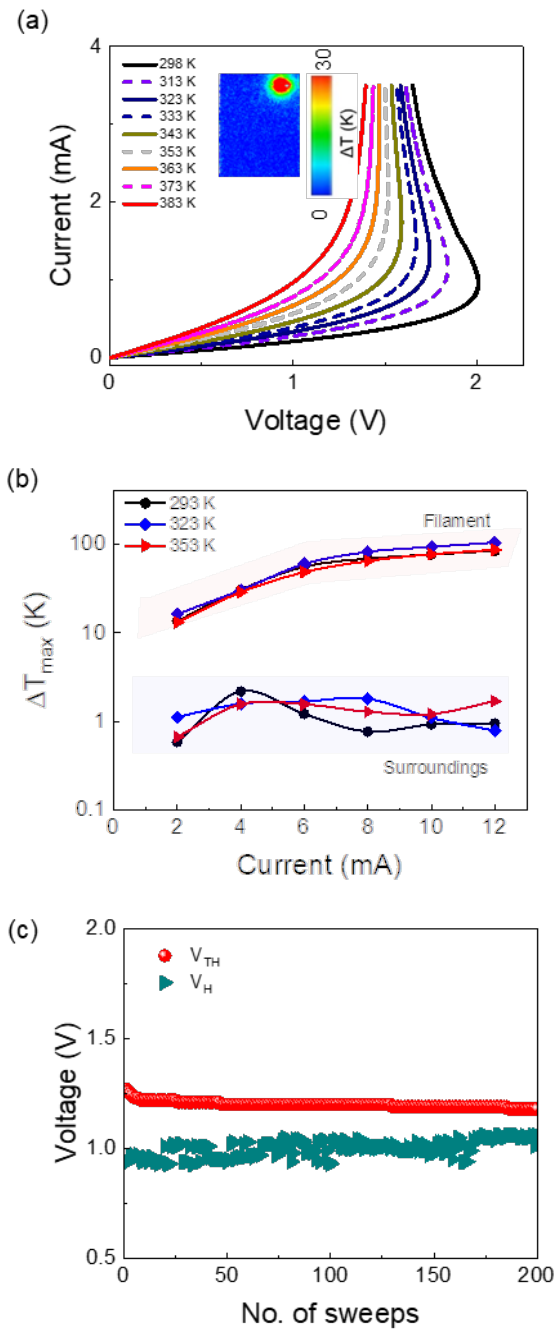

**Figure S8:** (a) Post-forming NDR in a device with Pt/Nb/NbO$_x$/Pt structure as a function of stage temperature, inset shows a temperature map showing filamentary conduction path after electroforming (further in-situ thermal mapping of NbO$_x$ devices can be found in our earlier publications[32, 39]), (b) maximum temperature in the filament and surrounding as a function of applied current. (c) Cycle-to-cycle variability of threshold and hold voltages in a Pt/Nb/Nb$_2$O$_5$/Pt device with a 30 nm Nb electrode.